# Combining electric and magnetic static field for the tuning of the lifetime of zero energy Feshbach resonances:  Application to $^3$He+NH($^3\Sigma$) collisions

*T. Stoecklin*
*Institut des Sciences Moléculaires, CNRS-UMR 5255, 351 cours de la Libération 33405 Talence, France*

**Abstract**

We study the variation of the positions of two magnetically tuned Zero energy Feshbach resonances when a parallel superimposed electric field is applied. We show that their variation as a function of the electric field follows a simple analytical law and is then predictable. We find that depending on the initial state of the diatomic molecule the resonance is either shifted to higher or to lower values of the magnetic field when the electric field is applied. We calculate the Close Coupling lifetimes of these resonances and show that they also follow a simple law as a function of both the magnetic and the electric field. We demonstrate using this expression that the lifetime of the resonances can be maximised by choosing an appropriate value of the applied electric and found a good agreement with the results of our Close Coupling calculations. These results could be checked in future experiments dedicated to the $^3$He + NH($^3\Sigma$) collisions

In the presence of an applied magnetic field zero energy Feshbach resonances are currently used to produce condensates of diatomic molecules from an ultra cold gas of atoms [1,2]. Such resonances also appear in collisions between atoms and molecules in the presence of an applied magnetic field and could be potentially used to produce ultra cold complexes. On this account and because of their possible use in the optimization of the cooling processes and the trapping techniques many theoretical studies are dedicated to this subject [3,4,5,6,7]. The tuning of the Feshbach resonances using a magnetic field for atom diatom collisions was first considered by Volpi and Bohn in their study dedicated to He-$O_2$($^3\Sigma$) [8]. The formal theory of collisions between atom and molecule in the presence of a magnetic field was then developed by Krems and Dalgarno [9] and applied by them to the He+CaH($^2\Sigma^+$) and Ar+NH($^3\Sigma^-$) collisions. The He-NH($^3\Sigma^-$) collision was then the subject of three detailed studies [10,11,12]. Two zero energy Feshbach resonances were located by the team of Hutson [12] for this system as a function of the magnetic field and we will focus in the present work on the action of a superimposed electric field on these two resonances. We use the extension of the formalism of Krems and Dalgarno [9] developed by Tscherbul and Krems[13] to include the interaction with a superimposed parallel electric field. Within this approach we take into account the fine structure of the diatomic molecule but disregard the molecular hyperfine interaction. The theoretical evaluation of the lifetimes of the complexes produced in this way is valuable information for the experimentalists and the Smith lifetime matrix Q [14] is the tool of choice for this purpose. The subject of this paper is to demonstrate that the lifetime of this resonance can be tuned using a superimposed parallel electric field

This required only little modifications of the code we developed to treat the case of the He-$N_2^+$($^2\Sigma$) collisions [15,16,17] in the presence of a magnetic field [18] and we will not enter again into the details of the method. We will only remind the main steps of our calculations in order to define the notations used in this paper. This approach uses a primitive uncoupled basis set to describe the field dressed states of the free diatomic molecule. This is a simple basis set $\phi_i = \chi_{v,N}(r)|NM_N\rangle|SM_S\rangle$ made of the products of the eigenfunction of the rotational ($\vec{N}^2$, $N_Z$) and electronic spin ($\vec{s}^2$, $S_Z$) angular momenta of the diatomic molecule, where $N_Z$ and $S_Z$ designate the projection of the angular momenta $\vec{N}$ and $\vec{s}$ along the direction of the applied field. As the asymptotical hamiltonian (1) describing the free molecule in the presence of the fields includes the spin rotation and the spin spin interactions as well as the Zeeman and

Stark terms, it is not diagonal in the uncoupled basis set $\phi_i$ which then cannot be used to describe the asymptotic states of the field dressed molecule.

$$H_{Diatom}^{Asymp} = H_{Diatom}^{SpinFree} + \gamma \cdot \vec{S} + \frac{2}{3}\lambda_{SS}\sqrt{\frac{24\pi}{5}}\sum_q Y_2^{q*}\left[\vec{S} \otimes \vec{S}\right]_q^2 + g\mu_0 \vec{B} \cdot \vec{S} - \vec{E} \cdot \vec{d} \qquad (1)$$

where $\gamma$ and $\lambda_{SS}$ are the spin-rotation and the spin spin interaction constants, g is the g factor for the electron, $\mu_B$ is the Bohr magneton and $\vec{d}$ the dipole moment of the diatomic molecule. $\vec{B}$ and $\vec{E}$ are the magnetic and the parallel electric field defining the Z space fixed axis. One uses instead the basis set $\chi_\alpha = \sum_i C_{\alpha i}\phi_i$ which diagonalises this Hamiltonian

$$\left[CH_{Diatom}^{Asymp} C^{-1}\right]_{\alpha\beta} = \xi_\alpha \delta_{\alpha\beta} \qquad (2)$$

While the electric field creates superpositions of different N states, the magnetic field removes the degeneracy in $M_J = M_N + M_S$ and each energy level $\xi_\alpha$ of the dressed diatomic molecule in the presence of the superimposed fields is associated with a single value of $M_J$ denoted $M_J(\alpha)$. For a given value of the projection of the total angular momentum along the direction of the fields $M_T$ and for a given $M_J(\alpha)$, the projection of the relative angular momentum L along the direction of the field basis set is then simply $M_L = M_T - M_J(\alpha)$. The basis set describing the collision process is then obtained by adding the possible values of the quantum number L for each value of $\alpha$. This basis set is denoted by the quantum numbers $\alpha$ $M_L$ L. In this basis set, the close coupling equations which have to be solved (3) and the asymptotic boundary conditions which have to be imposed in order to obtain the scattering matrix, take the form demonstrated in reference [9].

$$\left[\frac{d^2}{dR^2} - \frac{L(L+1)}{R^2} + 2\mu[E - \xi_\alpha]\right]F_{\alpha,M_L(\alpha),L}(R) = 2\mu \sum_{\alpha',M_L'(\alpha'),L'} \left[C^T U C\right]_{\alpha,M_L(\alpha),L}^{\alpha',M_L'(\alpha'),L'} F_{\alpha',M_L'(\alpha'),L'}(R) \qquad (3)$$

This system of coupled equations has in principle to be solved for each possible value of the total angular momentum projection $M_T$ in order to obtain the transition cross sections (4).

$$\sigma_{\alpha \to \alpha'}(E_\alpha) = \frac{\pi}{k_\alpha^2} \sum_{M_T} \sum_{M_L} \sum_L \sum_{M_L'} \sum_{L'} \left|T_{\alpha,M_L(\alpha),L;\alpha',M_L'(\alpha'),L'}^{M_T}\right|^2 \qquad (4)$$

As we are only interested in zero energy Feshbach resonances in the present work, we will calculate the scattering matrix for the single value of the projection $M_T$ of the total angular momentum corresponding to the s wave in the incident channel. We will however not calculate the cross sections in order to determine the magnetic field strength associated with the resonances $B_{res}(\varepsilon)$ for a given value of the applied electric field $\varepsilon$. We will use instead the fact that the magnetically tuned zero energy Feshbach resonances are corresponding to the poles of the scattering length $a_{le}(B,\varepsilon)$ as a function of the applied magnetic field as expressed in the formula (5) commonly used in atomic scattering [19].

$$a_{le}(B,\varepsilon) = a_{le}^{bg}\left[1 - \frac{\delta_B}{B - B_{res}(\varepsilon)}\right] \quad (5)$$

Where $a_{le}^{bg}$ is the background scattering length and $\delta_B$ the width of the resonance. Once a resonance is located, its lifetime has to be determined in order to discuss its experimental achievability.

$$Q = -i\hbar S^+\left[\frac{\partial S}{\partial E}\right]_{E=E_r} = i\hbar S\left[\frac{\partial S^+}{\partial E}\right]_{E=E_r} = Q^+ \quad (6)$$

The Smith lifetime matrix Q [14] defined as a function of the scattering matrix in equation (6) is the tool of choice for this purpose. In a forthcoming paper [20] we present an analytical method of calculation of the Smith lifetime Q matrix [14]. In the same work we show by replacing the expression (5) in the definition of the smith lifetime matrix given in equation (6) that we obtain the following expression for vanishing collision energy:

$$\lim_{k \to 0}(\tau) = -\frac{2\mu}{k}\text{Re}\left[a_{le}^{bg}\left[1 - \frac{\delta_B}{B - B_{res}(\varepsilon)}\right] - k^2\left(a_{le}^{bg}\right)^3\left[1 - \frac{\delta_B}{B - B_{res}(\varepsilon)}\right]^2 + ...\right] \quad (7)$$

where μ is the relative mass of the system and k the wave vector. We now focus on the two Zero Energy Feshbach Resonances (ZEFR) located by the team of Hutson **[12]** for the two values of the magnetic field (ZEFR1 around 7500 Gauss and ZEFR2 around 15000 Gauss) in their work dedicated to the $^3$He-NH($^3\Sigma$) collision. We use the same potential energy surface [10] and the same parameters as they used to perform our Close Coupling calculations. The main difference between the two sets of calculations is that our code uses the Magnus propagator instead of the Log Derivative method for the team of Hutson [12].

The magnetically tuned ZEFR results from the coupling of the initial channel with a closed channel induced by the interaction potential. These closed channels are the bound states of the He-NH complex that correlate to the upper levels of the field dressed diatomic molecule. When the electric field is applied, each level of the field dressed $^3\Sigma$ diatomic molecule is submitted to a second order Stark effect which results in quadratic shifts of the molecular energy levels as illustrated in **Figures** 1a and 1b respectively for the two values of the magnetic field (7500 and 15000 Gauss) associated with the two ZEFR identified by

Huston et al [12] for the $^3$He + NH($^3\Sigma$) collision. As can be seen on these figures, the energy shift $\Delta E_i$ of each individual level i takes the usual quadratic Stark form $\Delta E_i = a_i \varepsilon^2$ as a function of the electric field strength $\varepsilon$ but there is no significant Stark mixing of the rotational levels of NH for values of the electric field below 300 kV/cm as a result of the large value of the rotational constant of NH ($B_{rot}$=16.343 cm$^{-1}$) and of the relatively small value of its dipole moment ($\mu_D$=1.39 D).

As a consequence of the energy shifts of the field dressed levels of the diatomic molecule for a selected value $\varepsilon$ of the applied electric field, the value of the magnetic field which needs to be applied to obtain a given Zero Energy Feshbach Resonance (ZEFR) is also shifted from $\Delta B_{\text{Res}}(\varepsilon)$ when a superimposed parallel electric field is applied. If we denote in the absence of any applied electric field by $B_{\text{Res}}^0$ the value of the magnetic field associated with the position of the ZEFR and by $B_{\text{Res}}(\varepsilon)$ its value for a given applied electric field, we propose the Stark similar following form for the law of variation of the shift of a magnetically tuned zero energy Feshbach resonance as a function of the applied electric field

$$\frac{\Delta B_{\text{Res}}(\varepsilon)}{\varepsilon} = \frac{B_{\text{Res}}(\varepsilon) - B_{\text{Res}}^0}{\varepsilon} = a\varepsilon \qquad (8)$$

We reported respectively in **Figure** 2 a and 2b for both the ZEFR1 and ZEFR2 the result of our Close Coupling calculations for $\frac{\Delta B_{\text{Res}}(\varepsilon)}{\varepsilon}$ as a function of $\varepsilon$ for the fundamental state of the field dressed diatomic molecule ($\alpha$=1) and for the projection $M_T$=-1 of the total angular momentum. As can be seen on these figures, the proposed functional form (8) works quite well in the interval of electric field considered. We reported in **Table 1** the values of the least squares fit parameter a to the functional form given in equation (8) obtained from the Close Coupling calculations for the two ZEFR. The magnetic field strengths associated with these two ZEFR are both shifted to higher values when a superimposed parallel electric field is applied. However, the situation where the position of the ZEFR is shifted to lower values of the magnetic field when the electric field is applied can also be encountered as illustrated for the ZEFR2 in **Figure** 2b. This behaviour is obtained when performing calculations this time for the first excited level of the field dressed diatomic molecule ($\alpha$=2) and for $M_T$=0. The corresponding least squares fit value of the parameter a defined in equation (8) is also given in **Table 1**. Changing the initial state of the field dressed diatomic molecule from the fundamental to the first excited level changes the sign of the second order perturbation of this state due to the intermolecular potential. The absolute value of a is nearly the same showing that it results mainly from the coupling by the intermolecular potential between the states ($\alpha$=1) and ($\alpha$=2).

We now turn to an important issue when producing ultra cold complexes which is to determine their lifetimes. In a recent work [20], we took advantage of the simple analytical expression of the sector adiabatic wave functions of the Magnus propagator to obtain accurate values of the energy derivative of the S matrix which in turn is used to get the Smith lifetime Q matrix. Two examples of application of this method can be found in the two panels of the **Figures 3 and 4** where the minimum and the maximum eigenvalues of the Close Coupling Smith lifetime Q matrix are represented as a function of the magnetic field for the two ZEFR considered in this paper. **Figure** 3 is dedicated to the collision of the fundamental field dressed state ($\alpha$=1) of NH($^3\Sigma$) with $^3$He and for a projection $M_T$=-1 of the total angular momentum along the direction of the field while **Figure 4** was obtained for the collision of the field dressed state ($\alpha$=2) of NH($^3\Sigma$) with $^3$He and for a projection $M_T$=0 of the total angular momentum along the direction of the field. All the profiles of the curves obtained for other values of the applied electric field are similar for the two ZEFR. The singularity around the resonance which appears in equation (7) cannot be conveyed by a single eigenvalue of the Smith lifetime matrix and this is the reason why we reported both the highest and the lowest eigenvalue calculated. The Close Coupling Q matrix eigenvalues of both ZEFR follow the law of variation (7) but the singularity appears mainly in the positive eigenvalue in **Figure** 3 whereas it appears mainly in the negative eigenvalue in **Figure** 4. Another more important difference between the two ZEFR is shown in **Figures** 5 and 6 which are respectively dedicated to the ZEFR1 and ZEFR2. In **Figure** 5 and in the lower panel of **Figure** 6 we reported for different values of the applied electric field, the highest positive Close Coupling Q matrix eigenvalues as a function of the magnetic field for the collision of the fundamental field dressed state $\alpha$=1 of NH($^3\Sigma$) with $^3$He and for a projection $M_T$=-1 of the total angular momentum along the direction of the field. As can be seen on these figures the eigenlifetime decreases when the applied electric field increases. We obtain the opposite in the higher panel of **Figure** 6 where we reported for different values of the applied electric field, the highest positive Close Coupling Q matrix eigenvalues as a function of the magnetic field this time for the collision of the field dressed state ($\alpha$=2) of NH($^3\Sigma$) with $^3$He and for a projection $M_T$=0 of the total angular momentum along the direction of the field. As a matter of fact, in this case the eigenlifetime increases when the electric field increases. These different behaviors are predicted by the formula (9) which is obtained when replacing $B_{res}(\varepsilon)$ in equation (7) by the expression that we proposed in equation (8). We obtain the following expression for the lifetime as a function of both the applied magnetic and electric field:

$$\lim_{k \to 0}(\tau) = -\frac{2\mu}{k} \text{Re}\left[ a_{le}^{bg}\left[1 - \frac{\delta_B}{B - (B_{res}^0 + a\varepsilon^2)}\right] - k^2 (a_{le}^{bg})^3 \left[1 - \frac{\delta_B}{B - (B_{res}^0 + a\varepsilon^2)}\right]^2 + ... \right] \quad (9)$$

In this expression we did not mention the electric field dependence of both the background scattering length and the width $\delta B$ as it is assumed to be very feeble. We obtain in this case for the electric field derivative of this expression:

$$\lim_{k \to 0}\left(\frac{d\tau}{d\varepsilon}\right) = -\frac{2\mu}{k} \text{Re}\left[ \frac{a_{le}^{bg}(2a\varepsilon)\delta_B}{[B - (B_{res}^0 + a\varepsilon^2)]^2} \right] \quad (10)$$

This expression demonstrates that the lifetime has an extremum as a function of the electric field at zero field. If $a_{le}^{bg}$ is positive, this extremum is a maximum if the sign of a is negative or a minimum if a is positive. In any case this expression shows that the application of superimposed parallel electric and magnetic fields allows tuning the lifetime of the ZEFR. This is effectively what we found when performing Close Coupling Calculations as illustrated in **Figures** 5 and 6. When the initial state of the field dressed diatomic molecule is the fundamental one ($\alpha$=1) we see in **Table 1** that the sign of a is positive for both the ZEFR1 and ZEFR2. As the background scattering length is also positive in both cases the eigenlifteimes are maximum at zero field and decrease when the electric field strength increases as can be seen in **Figure** 5 and in the lower panel of **Figure** 6. Conversely for the first excited state ($\alpha$=2) of field dressed NH we see in **Table 1** that the sign of a is negative for the ZEFR2. As the background scattering length again is positive, the value of the eigenlifetime at zero field is a minimum and the eigenliftime increases when the applied electric field strength increases. The observed small departures from these general rules may be due to several factors. One of them is the possibly insufficient thinness of the magnetic field grid. It is also important to keep in mind that we neglected the field derivative of the background scattering length and of the resonance width in our demonstration. The formula (10) is obtained for a one dimensional Q matrix but in some cases there may be more than one highest non zero eigenvalue. These results will have in any case to be checked in future experiments but may be generalized to any molecule submitted to a quadratic Stark effect.

**Acknowledgments:**

I am very grateful to H. Cybulski, J. Klos, G. C. Groenenboom , A. Van der Avoid, D. Zgid, and G. Chalasiński for kindly providing me the potential energy surface of the He-NH($^3\Sigma$) collisions which they developed and is used in this paper.

**Table 1**: Quadratic Stark shift parameter a as defined in equation (8) obtained from a least squares fit of the Close Coupling results for the two zero energy Feshach resonances ZEFR1 and ZEFR2 and for the two initial states $\alpha=1$ and $\alpha=2$ of the field dressed diatomic molecule.

|   | ZEFR1 $\alpha=1$ | ZEFR2 $\alpha=1$ | ZEFR2 $\alpha=2$ |
|---|---|---|---|
| **a** | $8.38 \; 10^{-5}$ | $2.23 \; 10^{-3}$ | $-2.31 \; 10^{-3}$ |

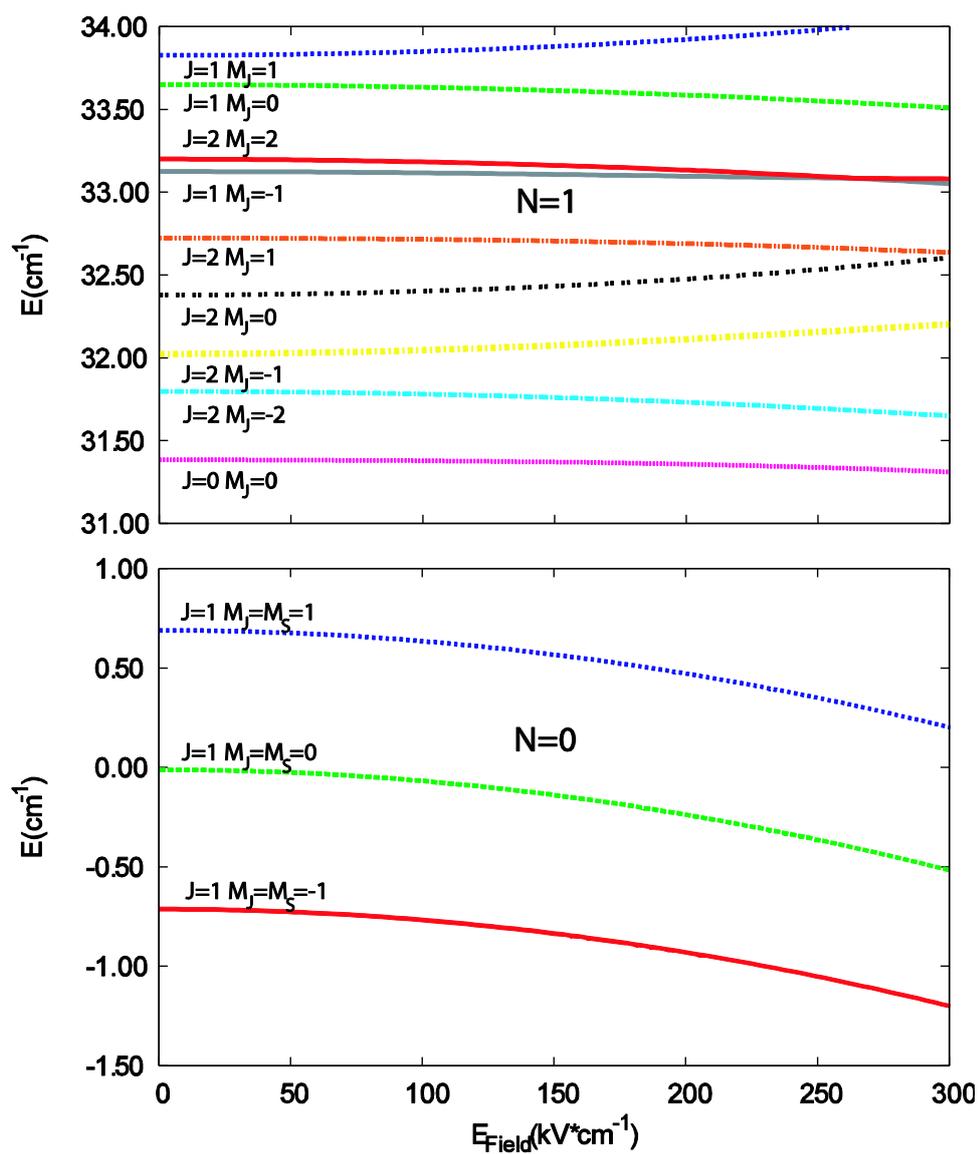

**Figure1a**: (Color online) Stark shifts of the NH($^3\Sigma$) molecule at a magnetic field of 7500 Gauss. The energy reference is referred to the ground rovibrational state of the molecule at zero field.

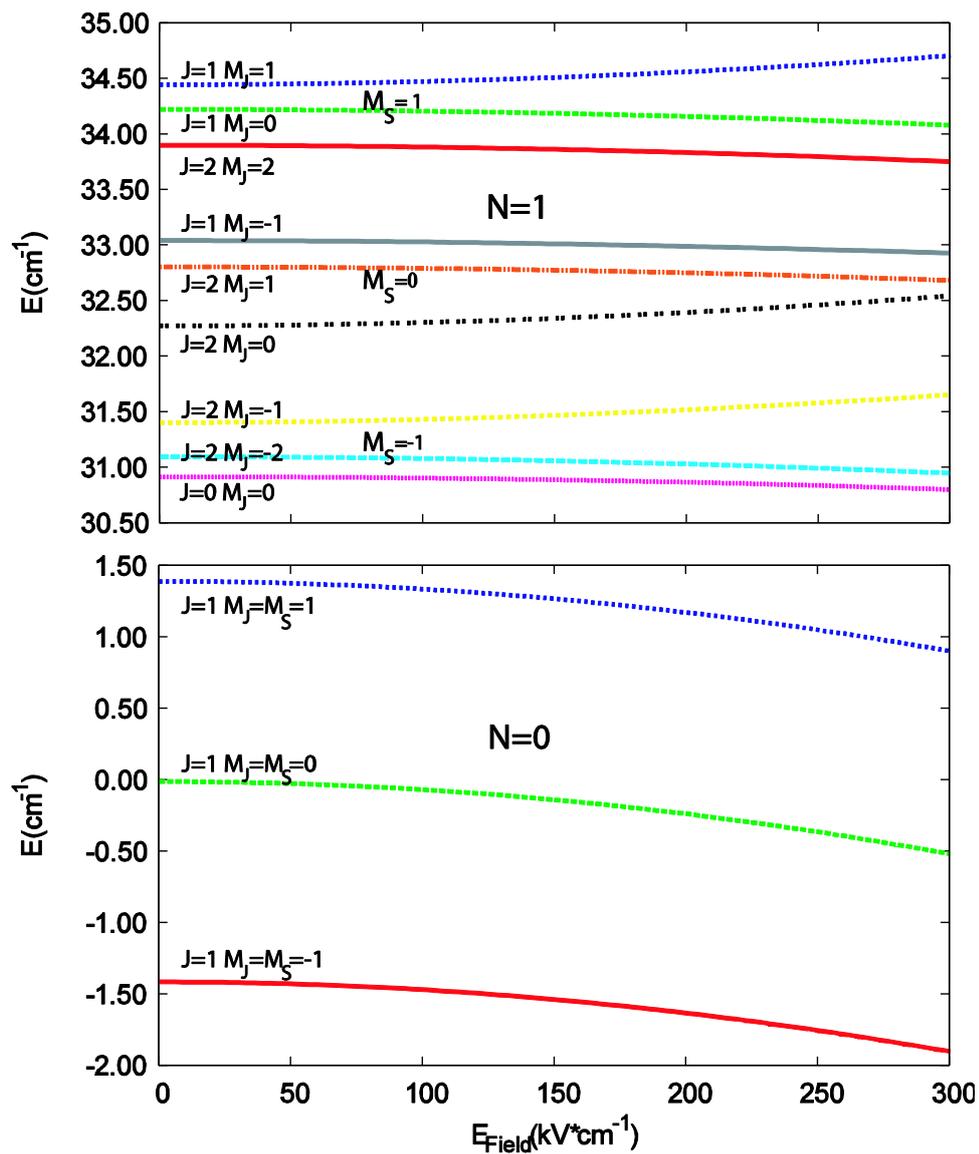

**Figure1b**: (Color online) Stark shifts of the NH($^3\Sigma$) molecule at a magnetic field of 15000 Gauss. The energy reference is referred to the ground rovibrational state of the molecule at zero field.

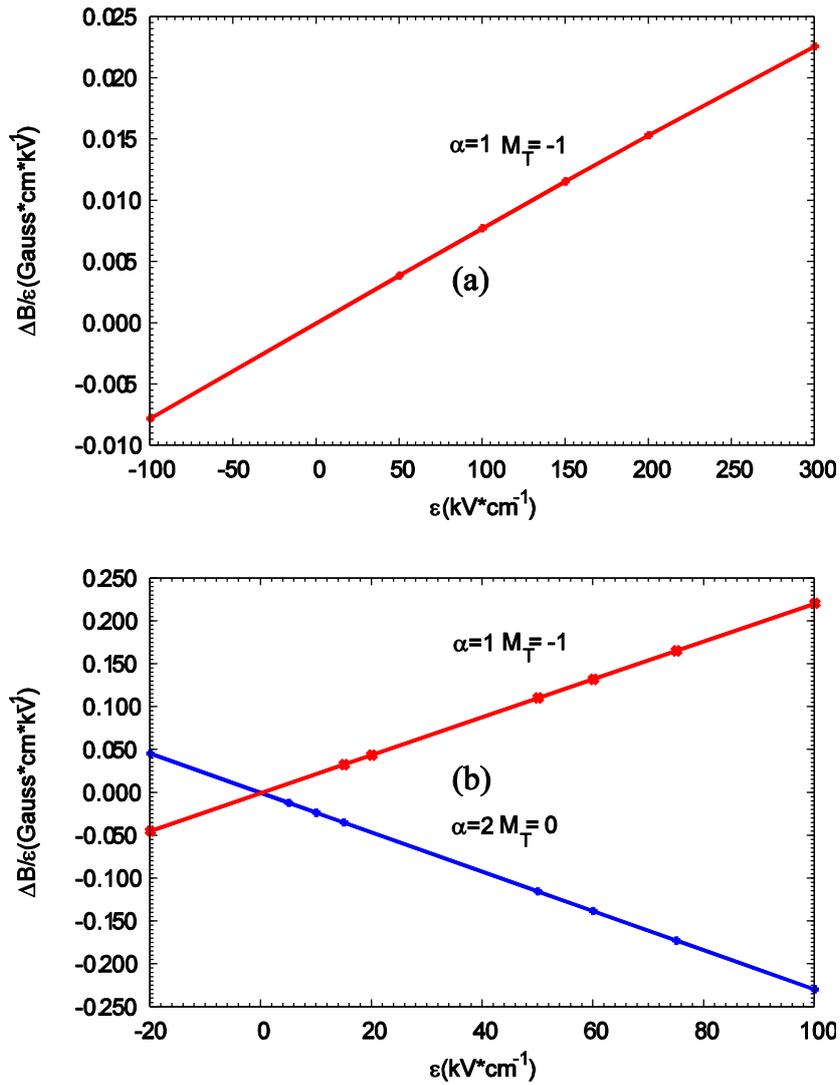

**Figure2**: (Color online) Close Coupling Stark shift divided by the electric field strength of the positions of the two magnetically tuned zero energy Feshbach resonance ZEFR1 and ZEFR2 as a function of the electric field strength for the $^3$He-NH($^3\Sigma$) collisions. The panels (a) and (b) are respectively dedicated to the ZEFR1 and ZEFR2. The value of the projection $M_T$ of the total angular momentum on the space fixed Z axis and the initial state of the field dressed diatomic molecule $\alpha$ are indicated on each curve.

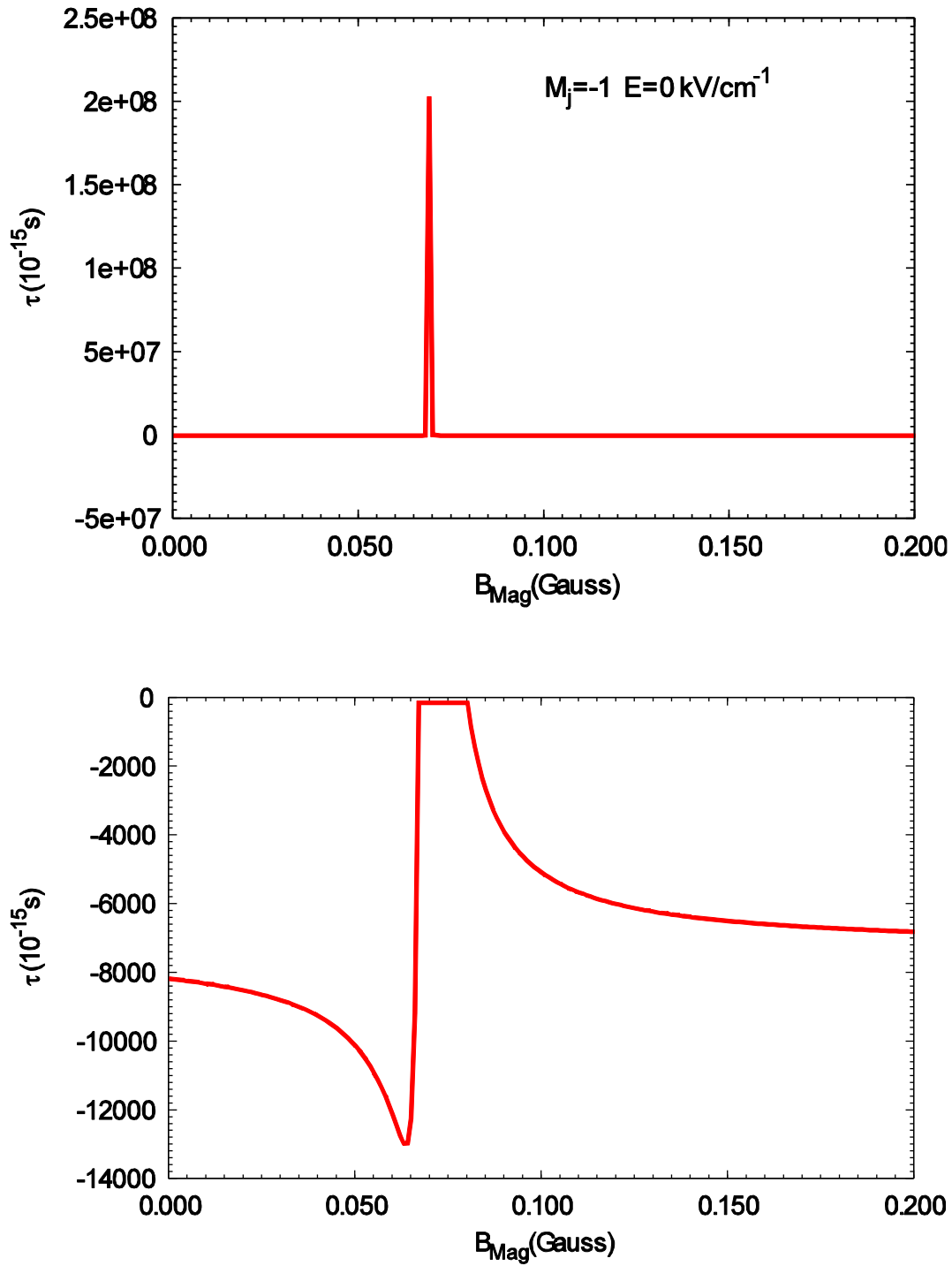

**Figure3**: (Color online) The lower and the higher panels show respectively the Close Coupling lowest negative and highest positive Q matrix eigenvalues as a function of the magnetic field for the collision of the fundamental field dressed state $\alpha=1$ of NH($^3\Sigma$) with $^3$He and for a projection $M_T=-1$ of the total angular momentum along the direction of the field. There is no electric field applied and the value of the magnetic field is in Gauss 7139.5 plus the value given on the abscissa axis.

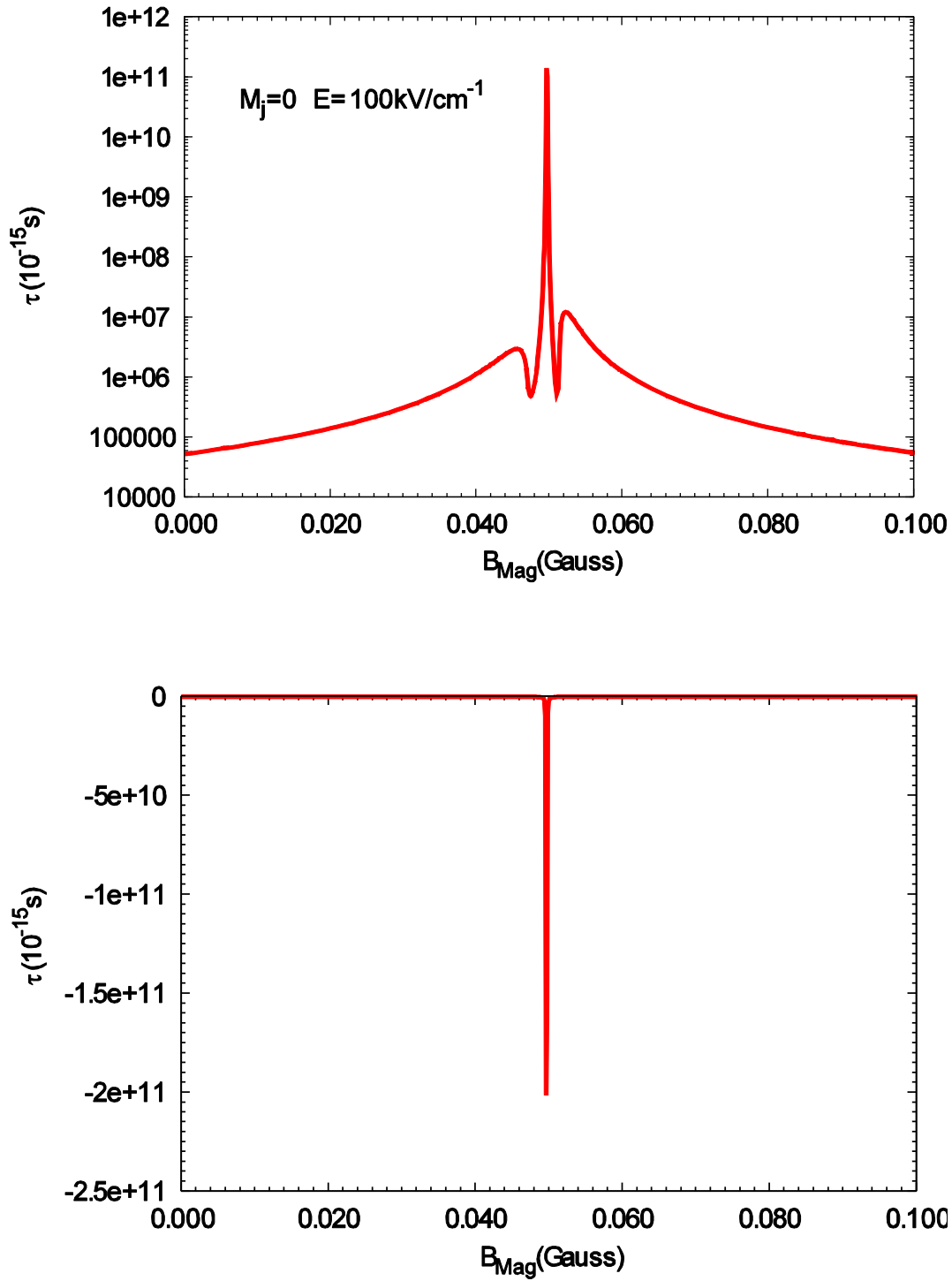

**Figure 4**: (Color online) The lower and the higher panels show respectively the Close Coupling lowest negative and highest positive Q matrix eigenvalues as a function of the magnetic field for the collision of the field dressed state $\alpha=2$ of NH($^3\Sigma$) with $^3$He and for a projection $M_T=0$ of the total angular momentum along the direction of the field. The electric field applied is 100 kV/cm and the value of the magnetic field is in Gauss 14259.03 plus the value given on the abscissa axis.

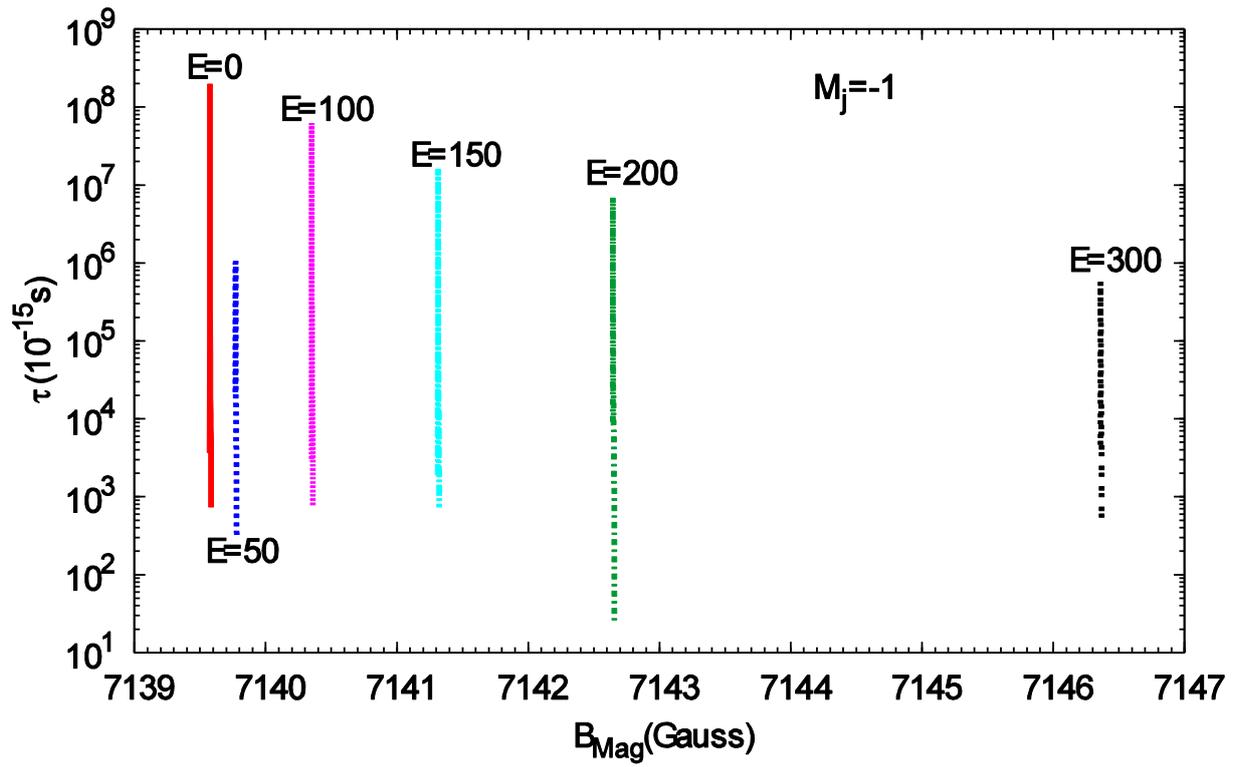

**Figure 5**: (Color online) Highest positive Close Coupling Q matrix eigenvalues as a function of the magnetic field for the collision of the fundamental field dressed state $\alpha=1$ of NH($^3\Sigma$) with $^3$He and for a projection $M_T=-1$ of the total angular momentum along the direction of the field. The value of the applied electric field is indicated on each curve and is given in kV/cm.

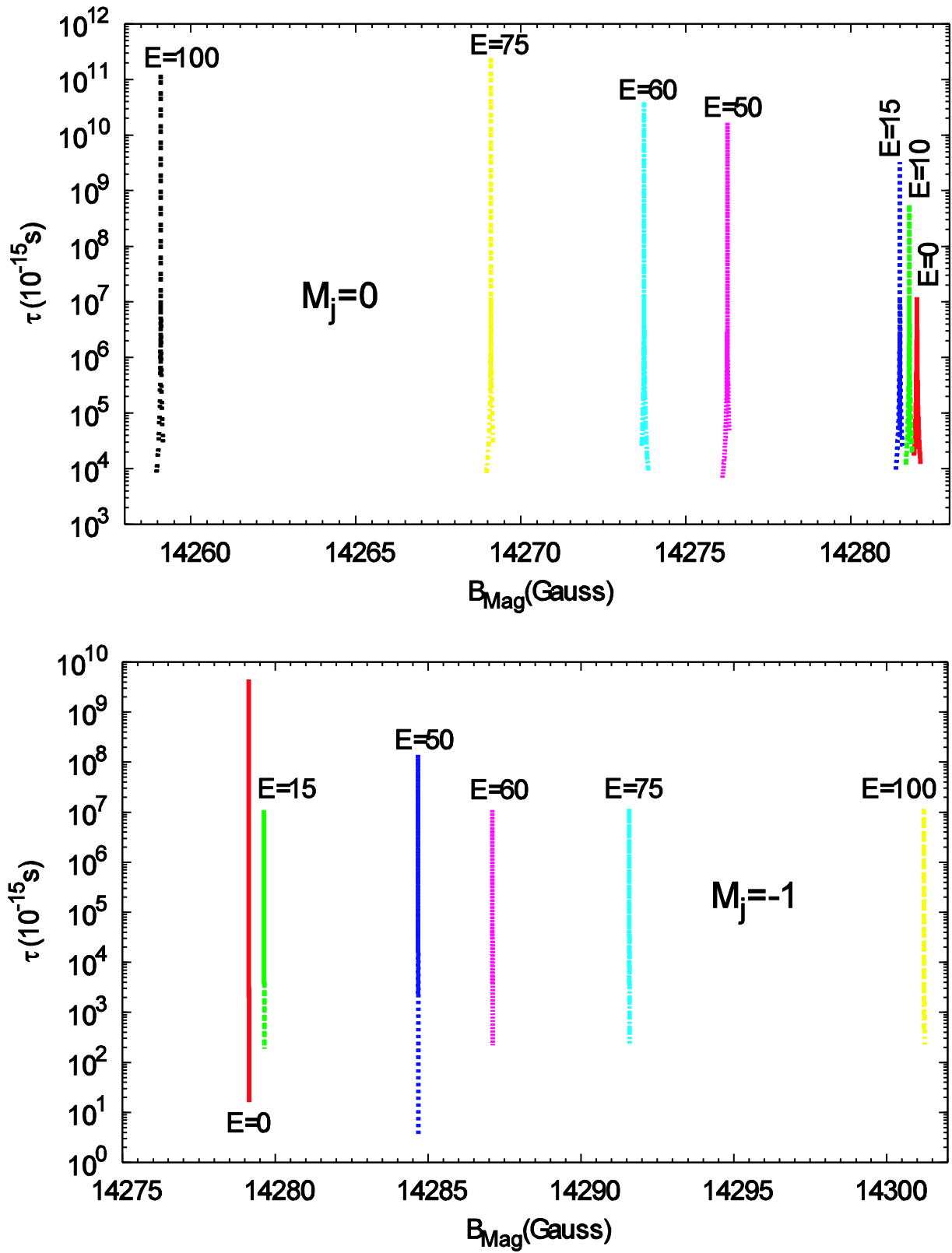

**Figure 6**: (Color online) Highest positive Close Coupling Q matrix eigenvalues as a function of the magnetic field for the collision of $^3$He with respectively NH($\alpha=1$) and $M_T=-1$ on the lower panel and NH($\alpha=2$) and $M_T=0$ on the higher panel. The value of the applied electric field is indicated on each curve and is given in kV/cm.